\documentclass[aps,prb,superscriptaddress,amsfonts,amsmath,amssymb,twocolumn,showpacs,floatfix]{revtex4}
\usepackage{url}
\usepackage{bm}
\usepackage{graphicx}
\usepackage{amsmath}
\usepackage{amstext}
\usepackage{amssymb}
\usepackage{amsfonts}
\usepackage{amsbsy}
\usepackage{verbatim}
\usepackage{color}
\usepackage[colorlinks=true, urlcolor=blue, linkcolor=blue, citecolor=blue, pdftex]{hyperref}
\usepackage{multirow}

\usepackage{ifthen}
\newlength{\abstractdiagwidth}
\newlength{\specdiagwidth}
\newcommand{\specdiag}[2][]{\parbox[c]{#2\specdiagwidth}{\includegraphics*[width=#2\specdiagwidth]{#1}}}%
\newcommand{\kag}[1]%
{%
\ifthenelse{\equal{#1}{hex1}}{\specdiag[KagDiag_hex1]{154}}{}%
\ifthenelse{\equal{#1}{hex2}}{\specdiag[KagDiag_hex2]{154}}{}%
\ifthenelse{\equal{#1}{2_1}}{\specdiag[KagDiag_2_1]{154}}{}%
\ifthenelse{\equal{#1}{2_2}}{\specdiag[KagDiag_2_2]{154}}{}%
\ifthenelse{\equal{#1}{3_1}}{\specdiag[KagDiag_3_1]{204}}{}%
\ifthenelse{\equal{#1}{3_2}}{\specdiag[KagDiag_3_2]{179}}{}%
\ifthenelse{\equal{#1}{3_3}}{\specdiag[KagDiag_3_3]{154}}{}%
\ifthenelse{\equal{#1}{3_4}}{\specdiag[KagDiag_3_4]{204}}{}%
\ifthenelse{\equal{#1}{3_5}}{\specdiag[KagDiag_3_5]{179}}{}%
\ifthenelse{\equal{#1}{3_6}}{\specdiag[KagDiag_3_6]{154}}{}%
\ifthenelse{\equal{#1}{4_1}}{\specdiag[KagDiag_4_1]{254}}{}%
\ifthenelse{\equal{#1}{4_2}}{\specdiag[KagDiag_4_2]{204}}{}%
\ifthenelse{\equal{#1}{4_3}}{\specdiag[KagDiag_4_3]{216.5}}{}%
\ifthenelse{\equal{#1}{4_4}}{\specdiag[KagDiag_4_4]{191.5}}{}%
\ifthenelse{\equal{#1}{4_5}}{\specdiag[KagDiag_4_5]{254}}{}%
\ifthenelse{\equal{#1}{4_6}}{\specdiag[KagDiag_4_6]{204}}{}%
\ifthenelse{\equal{#1}{4_7}}{\specdiag[KagDiag_4_7]{216.5}}{}%
\ifthenelse{\equal{#1}{4_8}}{\specdiag[KagDiag_4_8]{191.5}}{}%
\ifthenelse{\equal{#1}{5_1}}{\specdiag[KagDiag_5_1]{254}}{}%
\ifthenelse{\equal{#1}{5_2}}{\specdiag[KagDiag_5_2]{279}}{}%
\ifthenelse{\equal{#1}{5_3}}{\specdiag[KagDiag_5_3]{254}}{}%
\ifthenelse{\equal{#1}{5_4}}{\specdiag[KagDiag_5_4]{266.5}}{}%
\ifthenelse{\equal{#1}{5_5}}{\specdiag[KagDiag_5_5]{229}}{}%
\ifthenelse{\equal{#1}{5_6}}{\specdiag[KagDiag_5_6]{191.5}}{}%
\ifthenelse{\equal{#1}{5_7}}{\specdiag[KagDiag_5_7]{179}}{}%
\ifthenelse{\equal{#1}{5_8}}{\specdiag[KagDiag_5_8]{204}}{}%
\ifthenelse{\equal{#1}{5_9}}{\specdiag[KagDiag_5_9]{254}}{}%
\ifthenelse{\equal{#1}{5_10}}{\specdiag[KagDiag_5_10]{279}}{}%
\ifthenelse{\equal{#1}{5_11}}{\specdiag[KagDiag_5_11]{254}}{}%
\ifthenelse{\equal{#1}{5_12}}{\specdiag[KagDiag_5_12]{229}}{}%
\ifthenelse{\equal{#1}{5_13}}{\specdiag[KagDiag_5_13]{191.5}}{}%
\ifthenelse{\equal{#1}{5_14}}{\specdiag[KagDiag_5_14]{179}}{}%
\ifthenelse{\equal{#1}{5_15}}{\specdiag[KagDiag_5_15]{204}}{}%
\ifthenelse{\equal{#1}{5_16}}{\specdiag[KagDiag_5_16]{266.5}}{}%
\ifthenelse{\equal{#1}{6_1}}{\specdiag[KagDiag_6_1]{229}}{}%
\ifthenelse{\equal{#1}{6_2}}{\specdiag[KagDiag_6_2]{229}}{}%
\ifthenelse{\equal{#1}{6_3}}{\specdiag[KagDiag_6_3]{241.5}}{}%
\ifthenelse{\equal{#1}{6_4}}{\specdiag[KagDiag_6_4]{279}}{}%
\ifthenelse{\equal{#1}{6_5}}{\specdiag[KagDiag_6_5]{279}}{}%
\ifthenelse{\equal{#1}{6_6}}{\specdiag[KagDiag_6_6]{229}}{}%
\ifthenelse{\equal{#1}{6_7}}{\specdiag[KagDiag_6_7]{241.5}}{}%
\ifthenelse{\equal{#1}{6_8}}{\specdiag[KagDiag_6_8]{279}}{}%
\ifthenelse{\equal{#1}{6_9}}{\specdiag[KagDiag_6_9]{279}}{}%
\ifthenelse{\equal{#1}{6_10}}{\specdiag[KagDiag_6_10]{229}}{}%
\ifthenelse{\equal{#1}{6_11}}{\specdiag[KagDiag_6_11]{216.5}}{}%
\ifthenelse{\equal{#1}{6_12}}{\specdiag[KagDiag_6_12]{254}}{}%
\ifthenelse{\equal{#1}{6_13}}{\specdiag[KagDiag_6_13]{266.5}}{}%
\ifthenelse{\equal{#1}{6_14}}{\specdiag[KagDiag_6_14]{266.5}}{}%
\ifthenelse{\equal{#1}{6_15}}{\specdiag[KagDiag_6_15]{216.5}}{}%
\ifthenelse{\equal{#1}{6_16}}{\specdiag[KagDiag_6_16]{254}}{}%
\ifthenelse{\equal{#1}{6_17}}{\specdiag[KagDiag_6_17]{254}}{}%
\ifthenelse{\equal{#1}{6_18}}{\specdiag[KagDiag_6_18]{254}}{}%
\ifthenelse{\equal{#1}{6_19}}{\specdiag[KagDiag_6_19]{216.5}}{}%
\ifthenelse{\equal{#1}{6_20}}{\specdiag[KagDiag_6_20]{254}}{}%
\ifthenelse{\equal{#1}{6_21}}{\specdiag[KagDiag_6_21]{266.5}}{}%
\ifthenelse{\equal{#1}{6_22}}{\specdiag[KagDiag_6_22]{266.5}}{}%
\ifthenelse{\equal{#1}{6_23}}{\specdiag[KagDiag_6_23]{216.5}}{}%
\ifthenelse{\equal{#1}{6_24}}{\specdiag[KagDiag_6_24]{254}}{}%
\ifthenelse{\equal{#1}{6_25}}{\specdiag[KagDiag_6_25]{254}}{}%
\ifthenelse{\equal{#1}{6_26}}{\specdiag[KagDiag_6_26]{254}}{}%
\ifthenelse{\equal{#1}{6_27}}{\specdiag[KagDiag_6_27]{241.5}}{}%
\ifthenelse{\equal{#1}{6_28}}{\specdiag[KagDiag_6_28]{241.5}}{}%
\ifthenelse{\equal{#1}{6_29}}{\specdiag[KagDiag_6_29]{354}}{}%
\ifthenelse{\equal{#1}{6_30}}{\specdiag[KagDiag_6_30]{304}}{}%
\ifthenelse{\equal{#1}{6_31}}{\specdiag[KagDiag_6_31]{329}}{}%
\ifthenelse{\equal{#1}{6_32}}{\specdiag[KagDiag_6_32]{304}}{}%
\ifthenelse{\equal{#1}{6_33}}{\specdiag[KagDiag_6_33]{279}}{}%
\ifthenelse{\equal{#1}{6_34}}{\specdiag[KagDiag_6_34]{254}}{}%
\ifthenelse{\equal{#1}{6_35}}{\specdiag[KagDiag_6_35]{254}}{}%
\ifthenelse{\equal{#1}{6_36}}{\specdiag[KagDiag_6_36]{304}}{}%
\ifthenelse{\equal{#1}{6_37}}{\specdiag[KagDiag_6_37]{316.5}}{}%
\ifthenelse{\equal{#1}{6_38}}{\specdiag[KagDiag_6_38]{291.5}}{}%
\ifthenelse{\equal{#1}{7_1}}{\specdiag[KagDiag_7_1]{241.5}}{}%
\ifthenelse{\equal{#1}{7_2}}{\specdiag[KagDiag_7_2]{279}}{}%
\ifthenelse{\equal{#1}{7_3}}{\specdiag[KagDiag_7_3]{291.5}}{}%
\ifthenelse{\equal{#1}{7_4}}{\specdiag[KagDiag_7_4]{291.5}}{}%
\ifthenelse{\equal{#1}{7_5}}{\specdiag[KagDiag_7_5]{241.5}}{}%
\ifthenelse{\equal{#1}{7_6}}{\specdiag[KagDiag_7_6]{279}}{}%
\ifthenelse{\equal{#1}{7_7}}{\specdiag[KagDiag_7_7]{279}}{}%
\ifthenelse{\equal{#1}{7_8}}{\specdiag[KagDiag_7_8]{279}}{}%
\ifthenelse{\equal{#1}{7_9}}{\specdiag[KagDiag_7_9]{291.5}}{}%
\ifthenelse{\equal{#1}{7_10}}{\specdiag[KagDiag_7_10]{291.5}}{}%
\ifthenelse{\equal{#1}{7_11}}{\specdiag[KagDiag_7_11]{241.5}}{}%
\ifthenelse{\equal{#1}{7_12}}{\specdiag[KagDiag_7_12]{279}}{}%
\ifthenelse{\equal{#1}{7_13}}{\specdiag[KagDiag_7_13]{279}}{}%
\ifthenelse{\equal{#1}{7_14}}{\specdiag[KagDiag_7_14]{279}}{}%
\ifthenelse{\equal{#1}{7_15}}{\specdiag[KagDiag_7_15]{291.5}}{}%
\ifthenelse{\equal{#1}{7_16}}{\specdiag[KagDiag_7_16]{291.5}}{}%
\ifthenelse{\equal{#1}{7_17}}{\specdiag[KagDiag_7_17]{291.5}}{}%
\ifthenelse{\equal{#1}{7_18}}{\specdiag[KagDiag_7_18]{279}}{}%
\ifthenelse{\equal{#1}{7_19}}{\specdiag[KagDiag_7_19]{291.5}}{}%
\ifthenelse{\equal{#1}{7_20}}{\specdiag[KagDiag_7_20]{291.5}}{}%
\ifthenelse{\equal{#1}{7_21}}{\specdiag[KagDiag_7_21]{291.5}}{}%
\ifthenelse{\equal{#1}{7_22}}{\specdiag[KagDiag_7_22]{279}}{}%
\ifthenelse{\equal{#1}{7_23}}{\specdiag[KagDiag_7_23]{266.5}}{}%
\ifthenelse{\equal{#1}{7_24}}{\specdiag[KagDiag_7_24]{266.5}}{}%
\ifthenelse{\equal{#1}{7_25}}{\specdiag[KagDiag_7_25]{266.5}}{}%
\ifthenelse{\equal{#1}{7_26}}{\specdiag[KagDiag_7_26]{266.5}}{}%
\ifthenelse{\equal{#1}{7_27}}{\specdiag[KagDiag_7_27]{266.5}}{}%
\ifthenelse{\equal{#1}{7_28}}{\specdiag[KagDiag_7_28]{266.5}}{}%
\ifthenelse{\equal{#1}{7_29}}{\specdiag[KagDiag_7_29]{254}}{}%
\ifthenelse{\equal{#1}{7_30}}{\specdiag[KagDiag_7_30]{254}}{}%
\ifthenelse{\equal{#1}{7_31}}{\specdiag[KagDiag_7_31]{279}}{}%
\ifthenelse{\equal{#1}{7_32}}{\specdiag[KagDiag_7_32]{279}}{}%
\ifthenelse{\equal{#1}{7_33}}{\specdiag[KagDiag_7_33]{279}}{}%
\ifthenelse{\equal{#1}{7_34}}{\specdiag[KagDiag_7_34]{254}}{}%
\ifthenelse{\equal{#1}{7_35}}{\specdiag[KagDiag_7_35]{254}}{}%
\ifthenelse{\equal{#1}{7_36}}{\specdiag[KagDiag_7_36]{354}}{}%
\ifthenelse{\equal{#1}{7_37}}{\specdiag[KagDiag_7_37]{379}}{}%
\ifthenelse{\equal{#1}{7_38}}{\specdiag[KagDiag_7_38]{354}}{}%
\ifthenelse{\equal{#1}{7_39}}{\specdiag[KagDiag_7_39]{266.5}}{}%
\ifthenelse{\equal{#1}{7_40}}{\specdiag[KagDiag_7_40]{254}}{}%
\ifthenelse{\equal{#1}{7_41}}{\specdiag[KagDiag_7_41]{266.5}}{}%
\ifthenelse{\equal{#1}{7_42}}{\specdiag[KagDiag_7_42]{254}}{}%
\ifthenelse{\equal{#1}{7_43}}{\specdiag[KagDiag_7_43]{279}}{}%
\ifthenelse{\equal{#1}{7_44}}{\specdiag[KagDiag_7_44]{354}}{}%
\ifthenelse{\equal{#1}{7_45}}{\specdiag[KagDiag_7_45]{304}}{}%
\ifthenelse{\equal{#1}{7_46}}{\specdiag[KagDiag_7_46]{304}}{}%
\ifthenelse{\equal{#1}{7_47}}{\specdiag[KagDiag_7_47]{366.5}}{}%
\ifthenelse{\equal{#1}{7_48}}{\specdiag[KagDiag_7_48]{316.5}}{}%
\ifthenelse{\equal{#1}{7_49}}{\specdiag[KagDiag_7_49]{316.5}}{}%
\ifthenelse{\equal{#1}{7_50}}{\specdiag[KagDiag_7_50]{341.5}}{}%
\ifthenelse{\equal{#1}{7_51}}{\specdiag[KagDiag_7_51]{291.5}}{}%
\ifthenelse{\equal{#1}{7_52}}{\specdiag[KagDiag_7_52]{291.5}}{}%
}%

\def\fileversion{v1.2a}
\def\filedate{2007/11/23}
\makeatletter

\newbox\swb@xone
\newbox\swb@xtwo
\newbox\swb@xthree
\newbox\swb@xfour
\newdimen\swdimen@ne
\newdimen\swdimentw@

\newcommand{\acontraction}[5][1ex]{%
  \mathchoice
    {\acontraction@\displaystyle{#2}{#3}{#4}{#5}{#1}}%
    {\acontraction@\textstyle{#2}{#3}{#4}{#5}{#1}}%
    {\acontraction@\scriptstyle{#2}{#3}{#4}{#5}{#1}}%
    {\acontraction@\scriptscriptstyle{#2}{#3}{#4}{#5}{#1}}}%
\newcommand{\acontraction@}[6]{%
  \setbox\swb@xone=\hbox{${}#1{}#2{}$}%
  \setbox\swb@xtwo=\hbox{${}#1{}#3{}$}%
  \setbox\swb@xthree=\hbox{${}#1{}#4{}$}%
  \setbox\swb@xfour=\hbox{${}#1{}#5{}$}%
  \swdimen@ne=\wd\swb@xtwo%
  \advance\swdimen@ne by \wd\swb@xfour%
  \divide\swdimen@ne by 2%
  \advance\swdimen@ne by \wd\swb@xthree%
  \vbox{%
    \hbox to 0pt{%
      \kern \wd\swb@xone%
      \kern 0.5\wd\swb@xtwo%
      \acontraction@@{\swdimen@ne}{#6}%
      \hss}%
    \vskip 0.5ex
    \vskip\ht\swb@xtwo}}

\newcommand{\acontraction@@}[3][0.05em]{%
  \hbox{%
    \vrule width #1 height 0pt depth #3%
    \vrule width #2 height 0pt depth #1%
    \vrule width #1 height 0pt depth #3%
    \relax}}

\newcommand{\bcontraction}[5][1ex]{%
  \mathchoice
    {\bcontraction@\displaystyle{#2}{#3}{#4}{#5}{#1}}%
    {\bcontraction@\textstyle{#2}{#3}{#4}{#5}{#1}}%
    {\bcontraction@\scriptstyle{#2}{#3}{#4}{#5}{#1}}%
    {\bcontraction@\scriptscriptstyle{#2}{#3}{#4}{#5}{#1}}}%
\newcommand{\bcontraction@}[6]{%
  \setbox\swb@xone=\hbox{${}#1{}#2{}$}%
  \setbox\swb@xtwo=\hbox{${}#1{}#3{}$}%
  \setbox\swb@xthree=\hbox{${}#1{}#4{}$}%
  \setbox\swb@xfour=\hbox{${}#1{}#5{}$}%
  \swdimen@ne=\wd\swb@xtwo%
  \advance\swdimen@ne by \wd\swb@xfour%
  \divide\swdimen@ne by 2%
  \advance\swdimen@ne by \wd\swb@xthree%
  \lower 0.5ex \vbox{%
    \hbox to 0pt{%
      \kern \wd\swb@xone%
      \kern 0.5\wd\swb@xtwo%
      \bcontraction@@{\swdimen@ne}{#6}%
      \hss}%
    }}

\newcommand{\bcontraction@@}[3][0.05em]{%
  \hbox{%
    \swdimentw@=#3
    \advance\swdimentw@ by -#1
    \vrule width #1 height 0pt depth #3%
    \lower\swdimentw@\hbox{\vrule width #2 height 0pt depth #1}%
    \vrule width #1 height 0pt depth #3%
    \relax}}

\makeatother

\begin{document}

\title{Competing Valence Bond Crystals in the 
Kagome Quantum Dimer Model}

\author{\href{http://www.lpt.ups-tlse.fr/spip.php?article32}{Didier Poilblanc}}
\affiliation{Laboratoire de Physique Th\'eorique UMR-5152, CNRS and 
Universit\'e de Toulouse, F-31062 France}
\author{Gr\'egoire Misguich}
\affiliation{Institut de Physique Th\'eorique,
CEA, IPhT, CNRS, URA 2306, F-91191 Gif-sur-Yvette, France.}

\date{\today}

\begin{abstract} The singlet dynamics which plays a major role in the physics of the spin-1/2
Quantum Heisenberg Antiferromagnet (QHAF) 
on the Kagome lattice can be approximately described by projecting onto the nearest-neighbor 
valence bond (NNVB) singlet subspace. We re-visit here the effective Quantum Dimer Model which originates from
the latter NNVB-projected Heisenberg model via a non-perturbative Rokhsar-Kivelson-like scheme. 
By using Lanczos exact diagonalisation on a 108-site cluster supplemented by a careful symmetry analysis,
it is shown that a previously-found 36-site Valence Bond Crystal (VBC) in fact competes with a new type of
12-site  "{\it resonating-columnar}" VBC. The exceptionally large degeneracy of the GS multiplets (144
on our 108-site cluster) might 
reflect the proximity of the $\mathbb{Z}_2$ dimer liquid. Interestingly, these two VBC "emerge" in
{\it different topological sectors}. Implications for the interpretation of numerical results on the QHAF are outlined.

\end{abstract}
\maketitle

\section{Introduction}

\begin{figure}[htb]
\includegraphics[width=0.9\columnwidth]{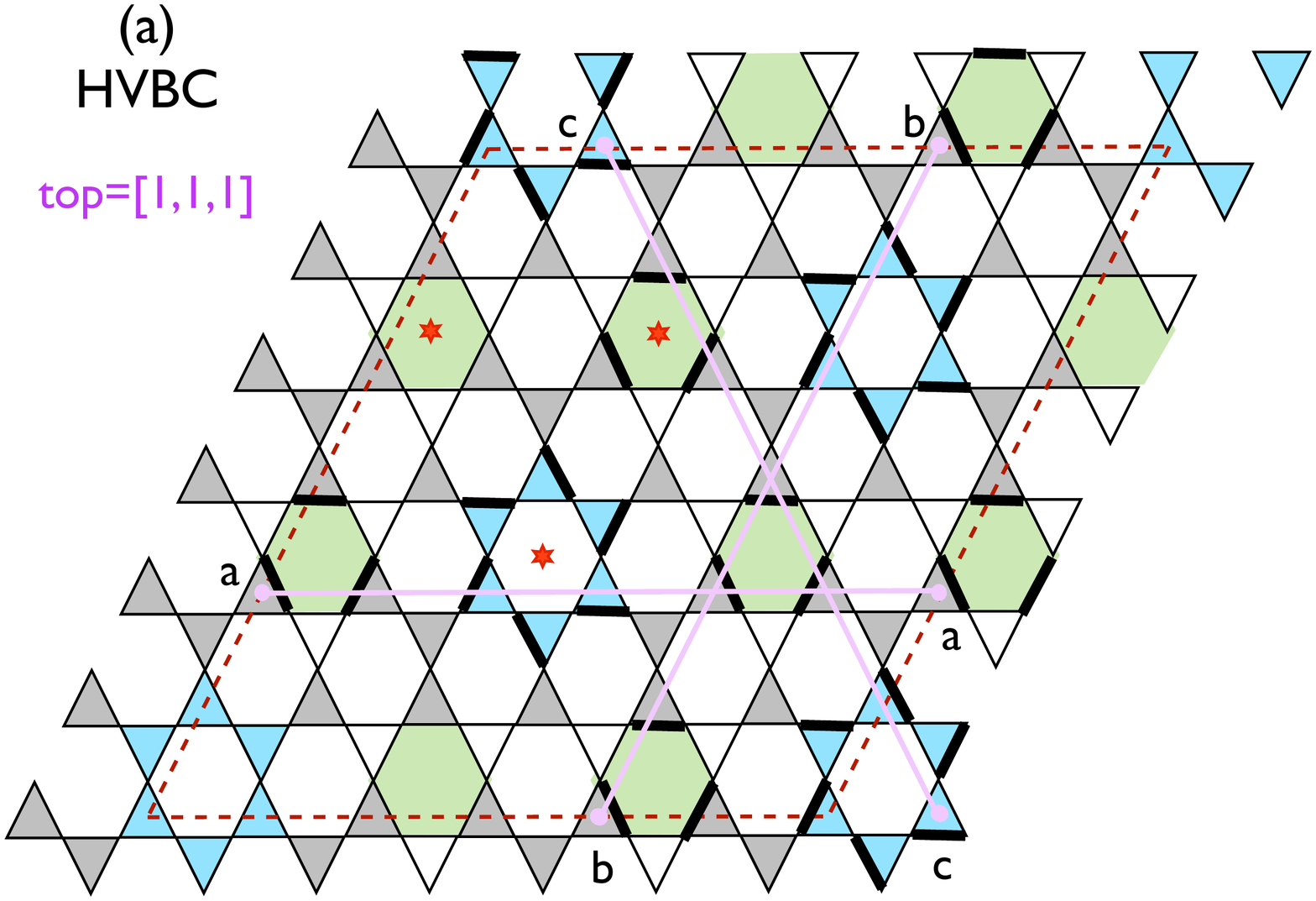}
\vskip -0.6cm
\includegraphics[width=0.9\columnwidth]{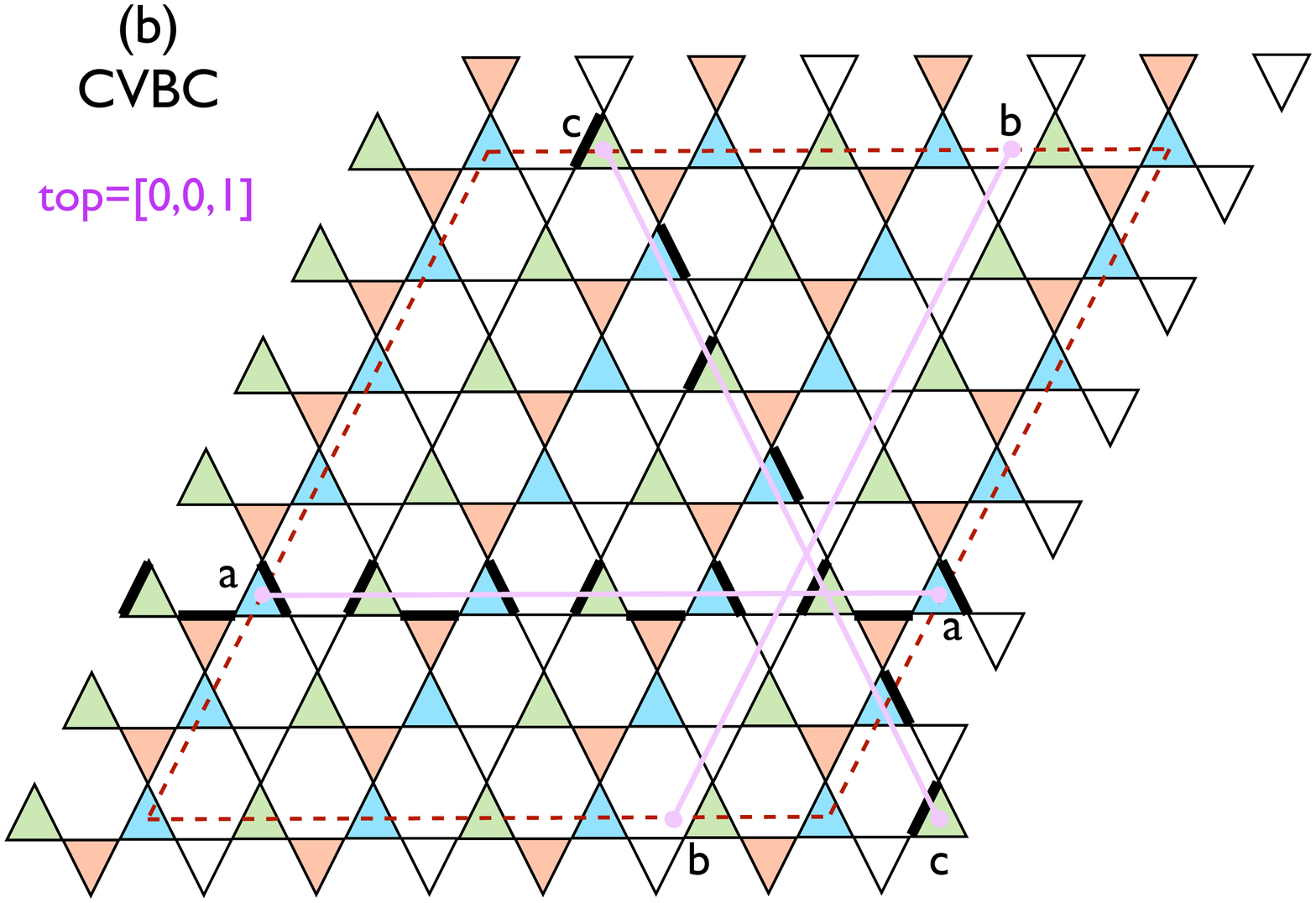}
\vskip -0.6cm
\caption{Kagome lattices. The 108-site clusters delimited by the (red) dashed lines are made of $6\times 6$ unit cells of 3-site up-triangles. Periodic boundaries are used. The four topological sectors are determined by the parities of the number of dimers (black thick bonds) cut by 
the three a-a b-b and c-c closed loops along the three lattice directions shown (see text). For clarity, the hard-core dimers are only shown along the three cuts. "Defect" triangles with no dimers are left empty.
(a) $2\sqrt{3}\times 2\sqrt{3} $ (36-site) HVBC showing resonating hexagons (shaded in green) and stars (shaded in blue)
in the [1,1,1] topological sector. (b) $2\times 1$ (6-site) CVBC with two types of up and two types of down triangles 
(all colored differently) in the [0,0,1] topological sector.
The red stars indicate the inversion ($\pi$-rotation) centers.
 }
\label{fig:kagome1}
\end{figure}

The spin-1/2 quantum Heisenberg antiferromagnet (QHAF) on the kagome lattice
(see Fig.~\ref{fig:kagome1}) is the paradigm of frustrated
quantum magnetism. Herbertsmithite~\cite{herbertsmithite} 
is an excellent experimental realization of such a kagome QHAF~: Copper atoms
carrying spin-1/2 degrees of freedom interacting antiferromagnetically are located on the
sites of perfect kagome layers.  In addition, deviations from an exact QHAF model like 
inter-layer couplings, Dzyaloshinski-Moriya magnetic interactions~\cite{dm},
intrinsic impurities~\cite{nmr}, etc... remain weak.  
The  {\it absence} of any magnetic ordering at the
lowest attainable temperatures~\cite{no-order} is the whole-mark of 
a non-magnetically ordered ground state as it may be realized in the ideal QHAF 
theoretical model (see below). However, it is still difficult to say whether the absence of a spin gap
as inferred from the low-temperature behavior of the susceptibility~\cite{review} can also be 
interpreted as a generic feature of the perfect (theoretical) QHAF or whether it is due to 
an extrinsic perturbation, as one of those mentioned above.

On the theoretical side, the proliferation of many low-energy singlets revealed in early 
Lanczos Exact Diagonalisations
(LED) of small clusters~\cite{LED} strongly suggests that a 
non-magnetically ordered phase could be stable in the Kagome QHAF. 
Spin liquids (SL)
preserving all symmetries, both spin-SU(2) and lattice translation, are the most fascinating and challenging candidates.
Algebraic SL~\cite{algebraic} or gapped SL~\cite{leung,dmrg} could be realized.
Recently,  Density Matrix Renormalization Group (DMRG) results~\cite{steve} on large Kagome strips 
found strong signals of a $\mathbb{Z}_2$ gapped SL.
However, in Variational Monte Carlo (VMC) studies~\cite{iqbal1} $Z_2$ SL (whose 
fermionic formulations were recently proposed~\cite{PALee}) were found to have slightly
higher energy than the algebraic SL.

\begin{figure}[htb]
\includegraphics[width=0.9\columnwidth]{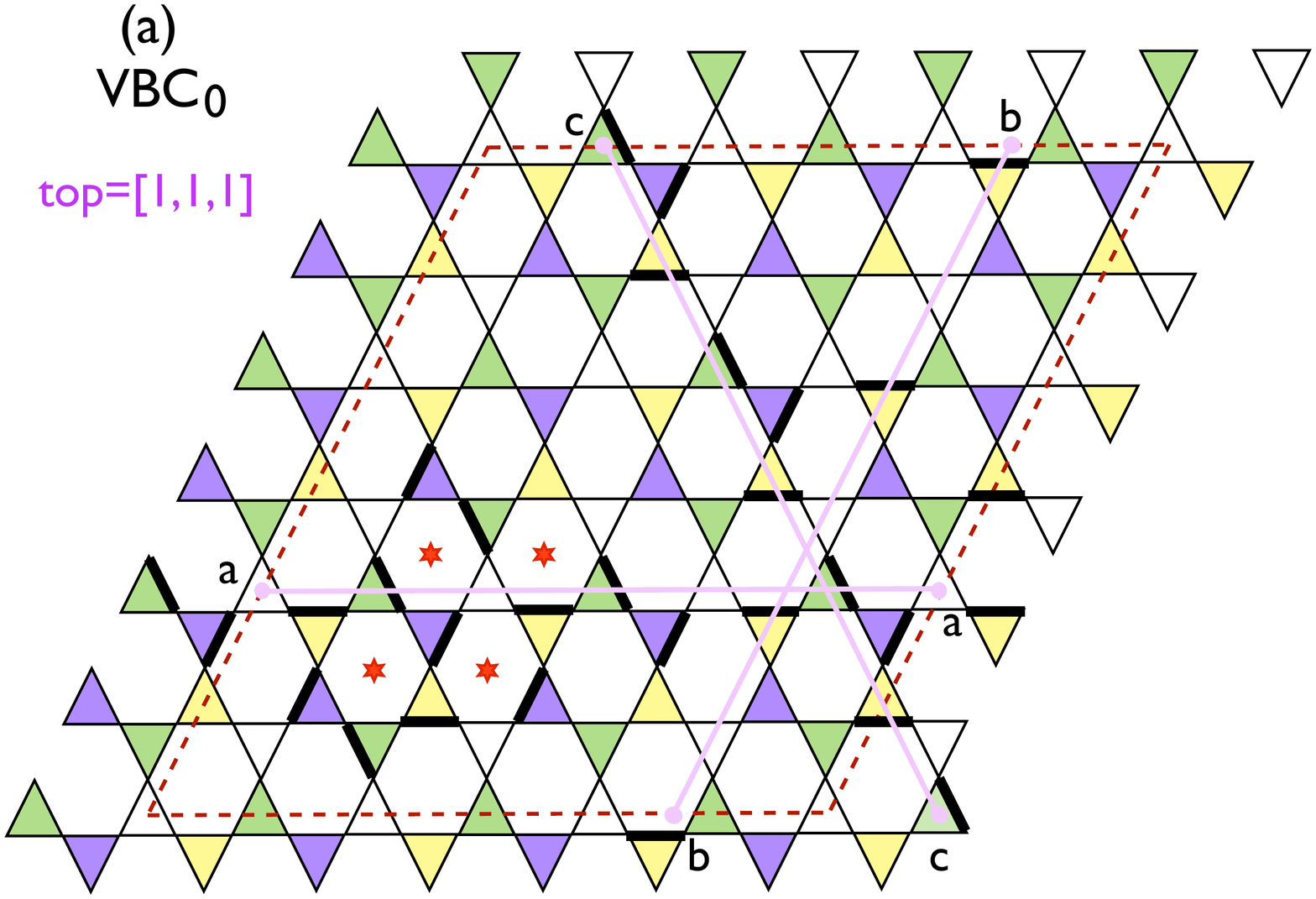}
\vskip -0.6cm
\includegraphics[width=0.9\columnwidth]{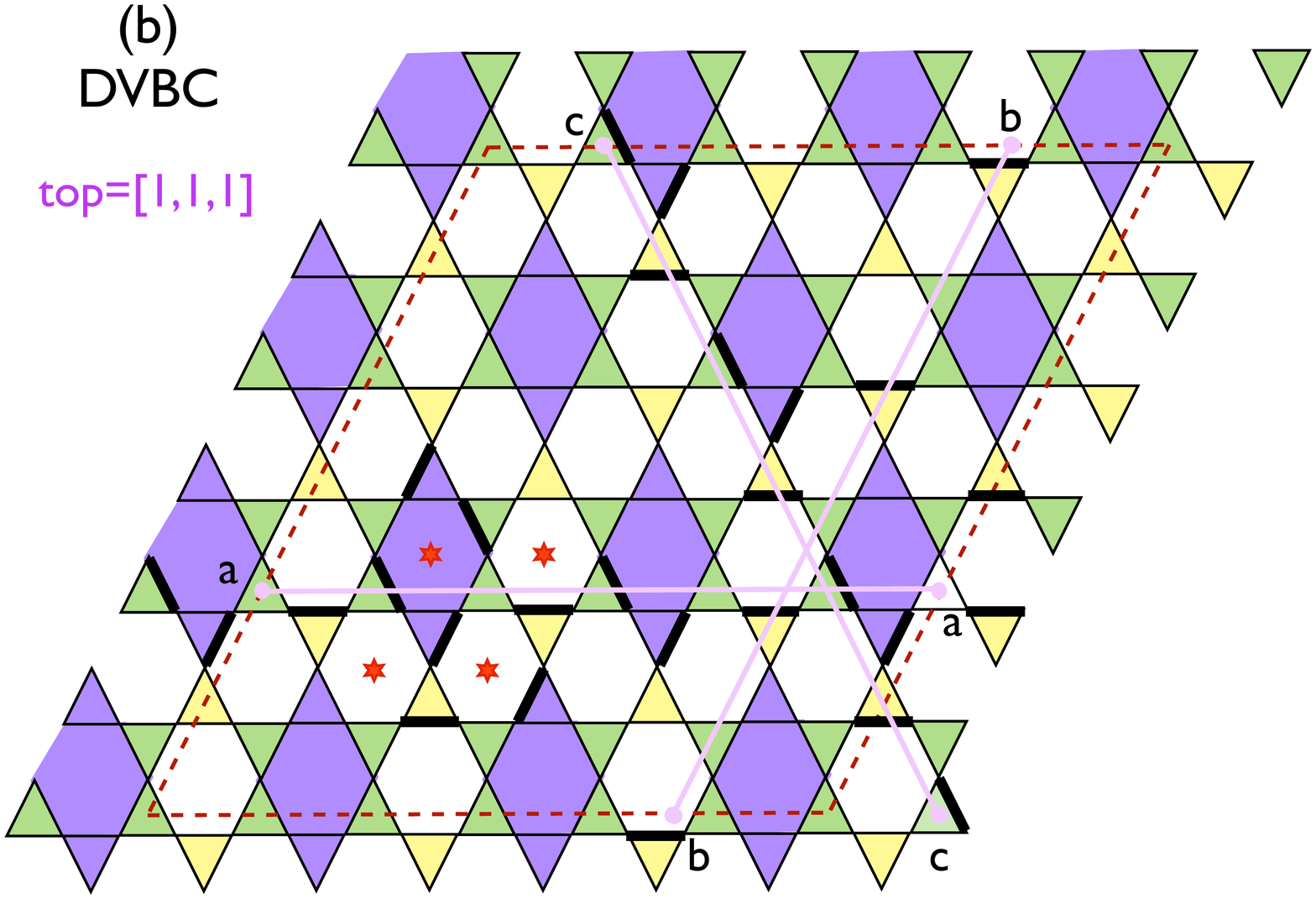}
\vskip -0.6cm
\includegraphics[width=0.9\columnwidth]{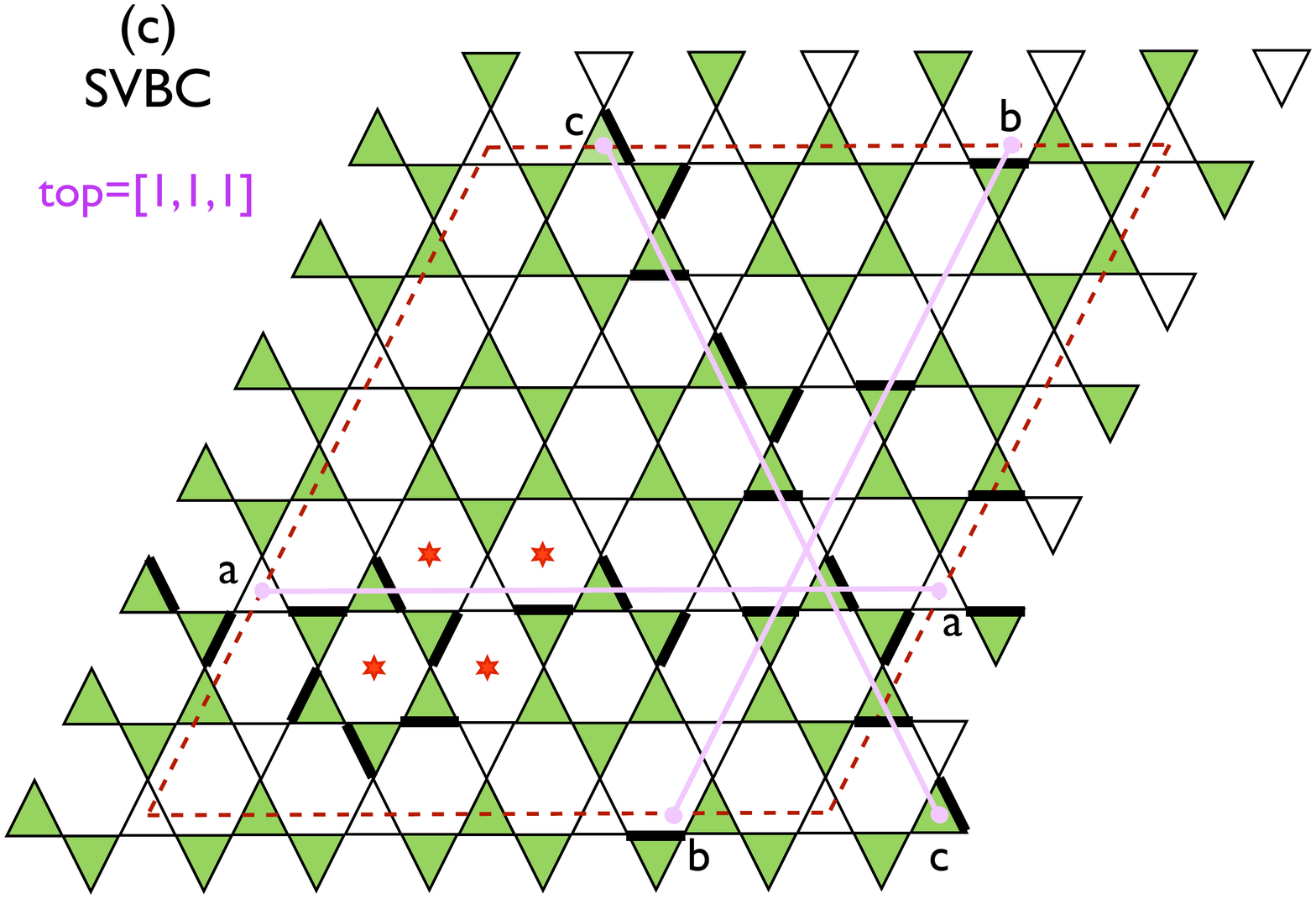}
\vskip -0.6cm
\caption{VBC with $2\times 2$ supercells in the [1,1,1] topological sector of the periodic $N=108$ site cluster.
(a) VBC$_0$ with 4 types of up (and down) triangles displayed with different colors except the
"defect" triangles left empty (degeneracy $g=8$). 
(b) The resonances of DVBC (colored diamonds) restore reflection symmetry w.r.t. the vertical and horizontal directions
but suppress the $\pi/3$-rotation symmetry ($g=12$).
(c) SVBC has {\it higher symmetry} than VBC$_0$, resonating stars restoring reflection symmetry w.r.t.  
the crystallographic axes ($g=4$). }
\label{fig:kagome2}
\end{figure}

Valence Bond Crystals (VBC) are other exotic candidates which, in contrast to SL, spontaneously 
break the elementary (3-site) unit cell translation symmetry, hence realizing a spontaneous (small) modulation characterized by a larger  unit cell (named here ``supercell'').
VBC with 12-site~\cite{sm02}, 18-site~\cite{marston} , 6-site~\cite{ba04} 
and 36-site~\cite{marston,senthil,singh} supercells have all been proposed 
in the literature as possible ground state (GS) of the Kagome QHAF. 
Very schematically, a VBC is often drawn as a frozen hardcore covering of the lattice
by nearest-neighbor spin singlets,  so called nearest neighbor valence bonds (NNVB). 
In reality, strong quantum fluctuations
are expected to severely reduce the dimer order parameter.
As shown in Fig.~\ref{fig:kagome1}(a), the 36-site VBC
shows an hexagonal lattice of resonating hexagons. It corresponds to a rather large
$2\sqrt{3}\times 2\sqrt{3}$ super-cell. The 6-site VBC exhibits  a $2\times 1$ super-cell with
columnar dimer order as shown in Fig.~\ref{fig:kagome1}(b). 
Both VBC are of particular interest in this study
and we shall refer to them as columnar VBC (CVBC) and hexagonal VBC (HVBC).
In VMC studies, a (fermionic) HVBC can be stabilized by a very small next NN {\it ferromagnetic}
exchange coupling.\cite{iqbal2}
The observation of "diamond" patterns characteristic of the diamond VBC (DVBC) displayed in Fig.~\ref{fig:kagome2}(b) 
(deriving from the "parent" VBC$_0$ of Fig.~\ref{fig:kagome2}(a)) was observed in recent DMRG~\cite{steve} studies.  
Also, closely related to VBC$_0$, the 12-site VBC of Ref.~\onlinecite{sm02} shown in Fig.~\ref{fig:kagome2}(c) 
displays a triangular lattice of resonating "stars" and shall be denominated as star VBC (SVBC).
 
 \section{Generalized quantum dimer model}
 
All these theoretical studies show that singlet dynamics plays a major role in the 
low-energy physics of the spin-1/2 Kagome QHAF.
One possible route is therefore to approximate the QHAF by projecting onto the NNVB
singlet {\it non-orthogonal} subspace.\cite{mambrini}
Starting from the NNVB-projected Heisenberg model, an effective Quantum Dimer Model (QDM) can be 
constructed via a non-perturbative Rokhsar-Kivelson (RK) like scheme.\cite{RK} 
This procedure bears the enormous advantage to provide a mapping to an {\it orthogonal} 
hard-core dimer basis. Details can be found in Refs.~\onlinecite{ralko,schwandt}. 
Due to the orthogonal nature of the dimer basis, the Kagome effective QDM can be diagonalized
(by the Lanczos algorithm) on clusters up to 108 sites ($6\times 6$ 3-site unit cells) providing 
clear signatures of the
HVBC and the proximity (in some parameter space) of a quantum transition to a $Z_2$ 
dimer liquid~\cite{poilblanc} (the hard-core dimer equivalent of the SL). Note that, strictly speaking, in 
the QDM framework the spinon (i.e. the elementary 
S=1/2 magnetic excitation) gap is pushed to infinity although some attempts have been carried out
to bring spinons back into the picture.\cite{spinons} 

A remarkable and useful feature of the effective QDM is that it provides simple and direct access to topological 
properties of the ground state, which is more involved in SU(2)-models.  On a periodic torus 
the parity of the numbers of dimers cut by two closed lines 
going along two orthogonal directions across the torus is conserved by the QDM Hamiltonian, hence defining 
four independent topological sectors. As clearly seen in a "toy" RK QDM defined in Ref.~\onlinecite{msp02}, 
the $\mathbb{Z}_2$ dimer liquid is characterized by {\it topological symmetry breaking} with 4 degenerate GS, one
in each topological sector. The $\mathbb{Z}_2$ SL also carries (gapped) excitations which are vortices 
of the dimer field (visons). In a cluster with an even (odd) number of sites, visons come in pairs~\cite{poilblanc} 
(in odd numbers~\cite{QDM-imp}). Interestingly, a class of dimer liquid-VBC quantum phase transition 
can be described by a specific field theory~\cite{sachdev} and has been observed numerically by 
interpolating the Hamiltonian between the toy RK QDM and the effective Kagome QDM.\cite{poilblanc}

In the present work, we shall use topological symmetry extensively to re-visit the effective QDM. 
From this refined analysis, we find that the previously-identified 36-site HVBC competes with new VBC.

We start here with the effective QDM derived in Ref.~\onlinecite{schwandt} (see Eq.~(42) of this paper),
\begin{widetext}
\setlength{\specdiagwidth}{0.06mm}
\begin{align}\label{eq:Hefffull}
H_{\rm eff} / J = &- \frac{4}{5}\kag{2_1} + \frac{1}{5}\kag{2_2} + \frac{16}{63}\left( \kag{3_1} + \kag{3_2} + \kag{3_3} \right) 
					+ \frac{2}{63} \left( \kag{3_4} + \kag{3_5} + \kag{3_6} \right)\\ 
					&- \frac{16}{255} \left( \kag{4_2} + \kag{4_3} + \kag{4_4} \right) 
	        + \frac{1}{255} \left( \kag{4_6} + \kag{4_7} + \kag{4_8} \right) + 0 \left( \kag{5_4} + \kag{5_16} \right)  \nonumber
\end{align}
\end{widetext}
The open diagrams correspond to kinetic processes around closed loops of length 6 (hexagons), 8, 10 or 12 (stars). In such a diagram, an alternating covering of (hard-core) dimers on the loop (i.e. a dimer on every second bond) is moved collectively by one step along the loop (i.e. it is "flipped"). Each diagram includes 
of course the reverse flipping action.
Also, by definition, each diagram is meant foanalysr all (non-equal)
loops of the lattices with the same topology i.e. obtained under all possible translations, reflections and
rotations (i.e. under all elements of the space group of the lattice). The second kind of diagrams (colored ones) are diagonal terms which simply "count" the number of "flippable" loops of the first kind. 
More details can be found e.g. in Refs.~\onlinecite{ralko,schwandt}.
Note that, in contrast to Ref.~\onlinecite{poilblanc}, Eq.~(\ref{eq:Hefffull}) includes infinite order
resummation of all the processes (loops) restricted to a single hexagon. 
This however only leads to very small changes of the values of the coefficients of $H_{\rm eff}$ 
compared to Ref.~\onlinecite{poilblanc}.  Note also that the stars have vanishing amplitudes at all
orders. 

\section{Topological properties and symmetry analysis}

The generalized QDM is investigated by Lanczos exact diagonalization of the  finite 108-site cluster
of the Kagome lattice made of $6\times 6$ unit cells of 3-site up-triangles and delimited by (red) 
dashed lines in Figs.~\ref{fig:kagome1},\ref{fig:kagome2}.
Periodic boundary conditions are used in the horizontal a-a and b-b (at 60$^o$--degrees) directions so that
the system is topologically equivalent to a torus, providing global invariance under the
translation group $\cal T$.  On such a geometry, the generalized QDM hamiltonian 
bears interesting topological properties: its Hilbert space of (orthogonal) NNVB hardcore coverings 
can be split in four disconnected topological sectors which are conserved by the QDM hamiltonian.
To determine the topological sector of a given dimer configuration, one simply needs to count the number 
of dimers (black thick bonds) cut by the three a-a b-b and c-c closed loops along the three lattice directions shown
in Figs.~\ref{fig:kagome1},\ref{fig:kagome2}. We shall label these sectors as [$\alpha$,$\beta$,$\gamma$]
where $\alpha$, $\beta$ and $\gamma$ are 0 or 1 depending whether the numbers of dimers 
cut by the corresponding loops are even or odd, respectively.
Every cluster of $P\times P$ unit cells (with $N=3P^2$ sites) contains a "symmetric" topological sector [$\alpha$,$\alpha$,$\alpha$]
with $\alpha=0$ or $\alpha=1$ for $P=4p$ or $P=4p+2$, respectively. Flipping the dimers along a close loop which goes around 
the torus, e.g. along the a-a direction, switches the parity of the number of dimers cut in the two other directions,
e.g. along b-b and c-c. For the 108-site cluster of interest here, one can then build the 4 topological sectors $T_0=[1,1,1]$,
$T_1=[1,0,0]$, $T_2=[0,1,0]$ and $T_3=[0,0,1]$. 
All configurations of each topological sector can be obtained separately by constructing an initial dimer configuration 
belonging to the targeted sector and then by applying recursively loop-flipping operators
of the type of Eq.~(\ref{eq:Hefffull}) to generate (numerically) new configurations until the Hilbert space is complete.

\begin{figure}[htb]
\includegraphics[width=0.9\columnwidth]{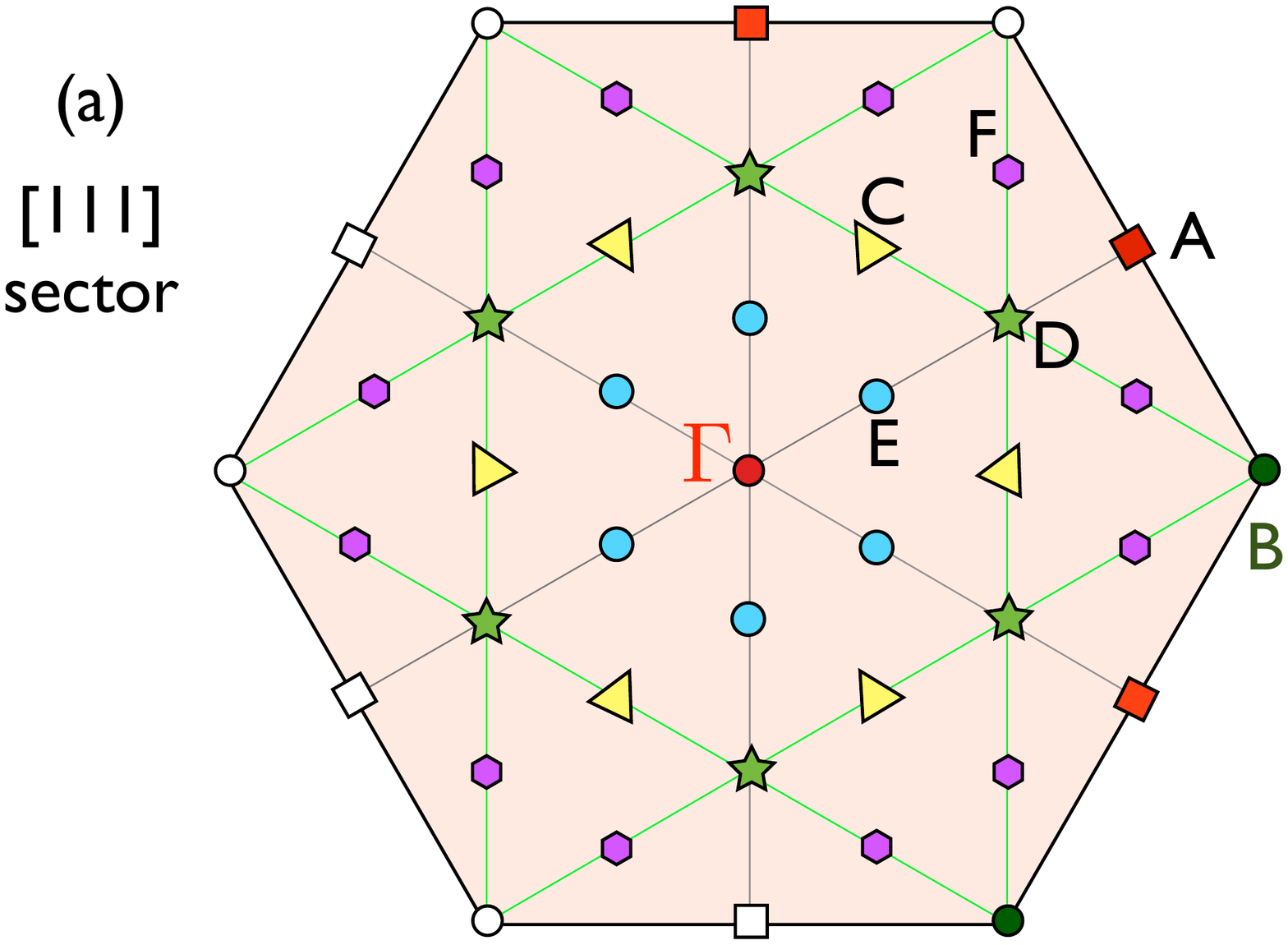}
\vskip -0.6cm
\includegraphics[width=0.9\columnwidth]{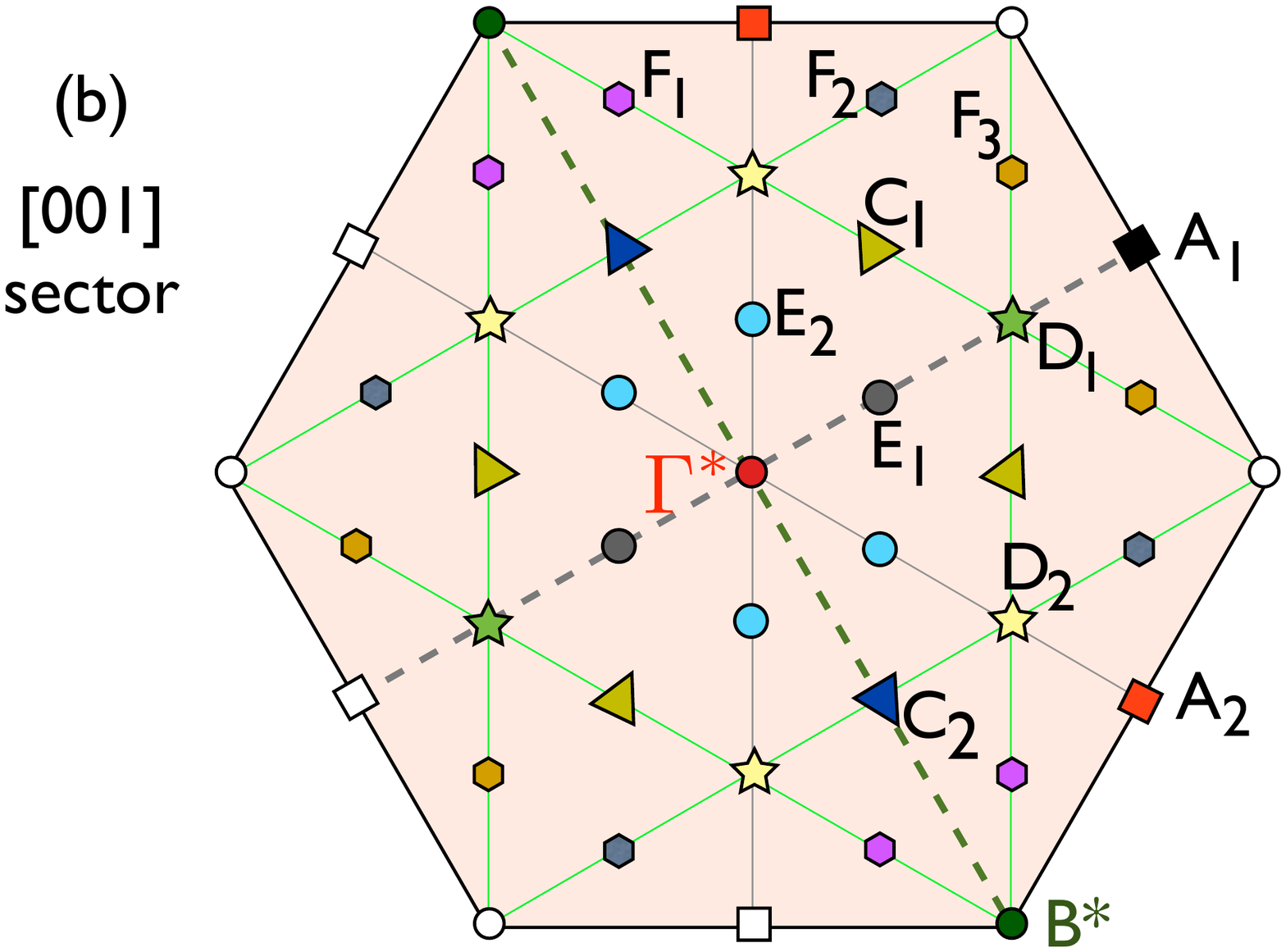}
\vskip -0.6cm
\caption{36 available momenta of the 108-site cluster ($6\times 6$ up-triangles) in the first 
Brillouin zone. Topological sectors have to be distinguished (as done on graph) because 
of their different point groups (see text). 
{\it Equivalent} momenta are represented by 
the same (filled) symbols. 
Translation by a reciprocal lattice vector (thin lines) of the momenta at the border of the 
1st BZ are shown by open symbols (correspond to the {\it same} momenta). 
The dashed lines in (b) represent the two (orthogonal) reflection symmetry axes of the $C_{2v}$ group. }
\label{fig:BZ}
\end{figure}

In order to block-diagonalize the Hamiltonian matrix in each topological sector, one can make use of all
space group symmetries leaving each sector globally invariant. 
The corresponding point group ${\cal G}(T_\alpha)$ depends on the topological sector $T_\alpha$. 
Note that we use here the center of the hexagon as the common origin for all point group elements. 
The full point group ${\cal G}=C_{6v}$ of the infinite lattice can be used 
for the "fully symmetric" $T_0=[1,1,1]$ topological sector.
In contrast,  the $2\pi/3$--rotation is lost in the $T_1=[1,0,0]$, $T_2=[0,1,0]$, and $T_3=[0,0,1]$ topological sectors so that the invariant point group is reduced to ${\cal G}=C_{2v}$. Each of these 3 topological sectors and their two corresponding (orthogonal) reflection symmetry axes are related 
by $2\pi/3$-- and $4\pi/3$--rotations.

\begin{table}[tbh]
 \begin{tabular}{@{} cccc|cccc @{}}
\toprule
$\bf K$ & $\cal G_{\bf K}$ & ${\rm card}\{\cal G_{\bf K}\}$ &  & &$\bf K$ & $\cal G_{\bf K}$  & ${\rm card}\{\cal G_{\bf K}\}$ \\
\hline 
$\Gamma$ & $C_{6v}$  &  12 & && $B^*$ &  $C_v$ & 2 \\
$A$ & $C_{2v}$  &  4 & && $C_1$ &  Id & 1 \\
$B$ & $C_{3v}$  &  6 && & $C_2$ &  $C_v$ & 2 \\
$C$ & $C_{v}$  &  2 && & $D_1$ &  $C_v$ & 2 \\
D& $C_{v}$  &  2 && & $D_2$ &  Id & 1 \\
E& $C_{v}$  &  2 & && $E_1$ &  $C_v$ & 2 \\
 F& Id  &  1 & && $E_2$ &  Id & 1 \\
$\Gamma^*$ & $C_{2v}$  &  4 & && $F_1$ &  Id & 1 \\
$A_1$ & $C_{2v}$  &  4 & && $F_2$ &  Id & 1 \\
$A_2$ & $C_{I}$  &  2 & && $F_3$ &  Id & 1 \\
\toprule
\end{tabular}
\caption{
Little groups $\cal G_{\bf K}$ for all $\bf K$ momenta of Fig.~\ref{fig:BZ}(a) and Fig.~\ref{fig:BZ}(b) corresponding to the $T_0=[1,1,1]$ and $T_3=[0,0,1]$ topological sectors, respectively.
}
\label{table:little_group}
\end{table}

In practice, one can specify a fixed momentum $\bf K$ in the Brillouin zones (BZ) of Fig.~\ref{fig:BZ}
and diagonalize the Hamiltonian separately in all irreducible representations (IRREP) of its little group $\cal G_{\bf K}$ (which is defined by all the elements of ${\cal G}(T_\alpha)$ leaving $\bf K$ invariant in the BZ).
All little groups are listed in Table~\ref{table:little_group} for the relevant $\bf K$-points of 
the $N=108$ cluster. 
The size of the Hilbert space (i.e. the $\#$ of orthogonal states) can then be reduced from the full
size $2^{N/3+1}\simeq 137\times10^9$ by (i) a factor 4 for each topological sector, 
(ii) a factor $N/3=36$
by using translation symmetry and (iii) a factor ${\rm card}\{\cal G_{\bf K}\}$ 
(see Table~\ref{table:little_group}) by using point symmetry. Hence, after proper construction of  the new "symmetrized" basis states in each IRREP, the linear sizes of the corresponding 
Hamiltonian "blocks" range from $\sim 80\times 10^6$ (${\cal G}_{\bf K}=C_{6v}$)
to $\sim 960\times 10^6$ (${\cal G}_{\bf K}={\rm Id}$). 
In the following, each IRREP will be labelled by its momentum/topological sector according to Fig.~\ref{fig:BZ}
(different combination of letters, subscripts and superscripts are used to distinguish all of them) 
and by the characters $(r_3,r_2,\sigma)$ 
which refer to $2\pi/3$ rotations, parity under inversion and 
reflection about the momentum direction, respectively,  whenever such symmetries are relevant 
(depends on ${\cal G}_{\bf K}$).
For the center of the BZ $\Gamma$, $\sigma$ refers to the reflection along one of the lattice directions
(consistently with Ref.~\onlinecite{sm07})
while, for the center $\Gamma^*$, it refers to the reflection along the axis perpendicular to the direction c 
(i.e. along the momentum $A_1$ in reciprocal space).

\section{Spontaneous symmetry breaking and degeneracies of VBC}

The spectrum of a system which spontaneously breaks some lattice symmetries at zero temperature
shows some ground-state degeneracies. Moreover, these degeneracies occur 
between levels belonging to different IRREP of the lattice symmetry group.
The simplest example is that of a $\mathbb{Z}_2$ symmetry breaking, where a state which in the ``even'' (topological)
sector is degenerate with a state in the ``odd'' sector.
In a finite size-system these degeneracies are only approximate, but they are
expected to become exponentially precise when increasing the system size. In turn, the presence of degeneracies is almost always the indication of some
spontaneous symmetry breaking, and this type of signature has been widely used in exact diagonalization studies to detect ordered phases.\cite{sm07}

If one knows the VBC pattern/symmetry, it is relatively easy to compute the IRREP which should appear
in the ground-state multiplet. Let us denote by
$|1\rangle,\cdots |d\rangle$ the states obtained by putting the VBC pattern on the lattice in all the possible positions/orientations.
For a sufficiently large lattice they can be considered as orthogonal to each other, and they transform into each other under the action of the lattice symmetry group $G={\cal G}\otimes  {\cal T}$:
for any symmetry $g\in G$ and for any state $|i\rangle$, there exist another state $j=f(g,i)$ such that $g|i\rangle=|j\rangle$.
We assume in particular that the later equation holds without any phase factor.
Such a ground-state multiplet
defines nothing but a (reducible) {\it representation} of dimension $d$ of the group $G$.  Determining the associated ground-state quantum numbers amounts to decompose this representation
over the IRREPs of $G$, which is a standard task in group theory. If we denote by $\rho$ some IRREP of $G$, the number $n_\rho$ of ground-states which should appear in the symmetry sector $\rho$
is given by a character formula:
\begin{equation}
  n_\rho=\frac{1}{|G|}\sum_{g\in G} \chi_\rho(g^{-1}) {\rm Tr}_{|1\rangle,\cdots |d\rangle}(g) \label{eq:mult}
\end{equation}
where $\chi_\rho(g^{-1})$ is the character of $\rho$ on the group element $g^{-1}$, $|G|$ is the total number of symmetries, and ${\rm Tr}_{|1\rangle,\cdots |d\rangle}(g)$
is the number of VBC states which are invariant under $g$: $g|i\rangle=|i\rangle$. The later quantity only depends on the symmetries of the VBC and not on the full wave-function. It can therefore be computed
using ``simplified'' states with the correct symmetries. It can sometimes be reduced to a {\it single} (non-resonant) covering, or a linear combination of $2^{n_{\rm cell}}$ coverings for resonating states with $n_{\rm cell}$ super cells ($n_{\rm cell}=N/{\cal N}$). In any case, from the group characters $\chi_\rho$ and the symmetries of the VBC, the
multiplicities $n_\rho$ can be systematically computed and compared to the spectrum obtained in LED. 

\begin{table}[htb]
\begin{center}
 \begin{tabular}{@{} ccccccc @{}}
   \toprule
VBC & cell & $\Gamma$ ($\times 1$)& A ($\times 3$)&  $\Gamma^*$ ($\times 1$)& A$_2$ ($\times 2$)& A$_1$ ($\times 1$)\\ 
   \hline
\footnotesize VBC$_0$ &$2\times 2$ & \scriptsize $(1,+,\pm)$ &  \scriptsize $(\times,+,\pm)$&   & & \\ 
 \footnotesize DVBC &$2\times 2$ & \scriptsize $(1,+,+)$ &  \scriptsize $(\times,+,+)$$^a$&   & & \\ 
 \footnotesize & & \scriptsize $(j,+,\times)$$^b$
 & \scriptsize $(\times,+,-)$ &   & & \\ 
\footnotesize SVBC &$2\times 2$ & \scriptsize $(1,+,+)$ &  \scriptsize $(\times,+,+)$&   & & \\ 
\footnotesize CVBC &$2\times 1$ & & & \scriptsize $(\times,\pm,\pm)$ &  \scriptsize $(\times,\pm,\times)$&    \\ 
\footnotesize $r{\rm VBC}_2$ &$2\times 2$ &&& \scriptsize $(\times,\pm,\pm)$ &  \scriptsize $(\times,\pm,\times)$$^a$
& \scriptsize $(\times,\pm,\pm)$  \\ 
\footnotesize $r{\rm VBC}_1$ &$2\times 2$ &&& \scriptsize $(\times,\pm,+)$ &  \scriptsize $(\times,\pm,\times)$& \scriptsize $(\times,\pm,+)$  \\ 
\footnotesize $r{\rm VBC}_1^\prime$ &$2\times 2$ &&& \scriptsize $(\times,+,+)$ &  \scriptsize $(\times,\pm,\times)$& \scriptsize $(\times,+,+)$  \\ 
 &&&& \scriptsize $(\times,-,-)$ &  & \scriptsize $(\times,-,-)$  \\ 
\footnotesize $r{\rm VBC}_3$ &$2\times 2$ &&& \scriptsize $(\times,+,\pm)$ &  \scriptsize $(\times,+,\times)$$^a$& \scriptsize $(\times,+,\pm)$  \\ 
\footnotesize VBC$^*$ &$2\times 2$ &&& \scriptsize $(\times,+,\pm)$ &  \scriptsize $(\times,+,\times)$$^a$& \scriptsize $(\times,+,\pm)$  \\ 
 \footnotesize $\mathbb{Z}_2$  & $1\times 1$ & \scriptsize $(1,+,+)$ &   &  \scriptsize $(\times,+,+)$  & & \\ 
     \toprule
 \end{tabular}
\end{center}
\caption{Quantum numbers of the GS multiplet for VBC with $m \times n$ super-cells
of ${\cal N}=3mn$ sites. 
The letters refer to the momenta of Fig.~\protect\ref{fig:BZ} (with their multiplicities in brackets) 
in the two topological sectors [1,1,1] and [0,0,1].
The character related to $2\pi/3$ rotations ($r_3=1,j,j^2$), parity under inversion ($r_2=+,-$) and 
reflection about the momentum direction ($\sigma=+,-$) are denoted as $(r_3,r_2,\sigma)$.   
``$\pm$" means 
that {\it both} even and odd GS
are present and ``$\times$'' stands for ``symmetry not relevant'' (e.g. $2\pi/3$ rotation
and hence $r_3$ is irrelevant in the [0,0,1] topological sector).
VBC$_0$, DVBC and SVBC are 8-fold, 12-fold and 4-fold degenerate, respectively. 
CVBC, $r{\rm VBC}_1$, $r{\rm VBC}_1^\prime$ and $r{\rm VBC}_3$
 have the same degeneracy $g=8$ in the [0,0,1] sector i.e. $g=8\times 3=24$ including all 3 topological sectors
 while $r{\rm VBC}_2$ has $g=16\times 3=48$.  The last line corresponds to 
 the $\mathbb{Z}_2$ dimer liquid ($g=4$).  }
\label{table:VBC}
\end{table}

\begin{figure}[htb]
\includegraphics[width=0.9\columnwidth]{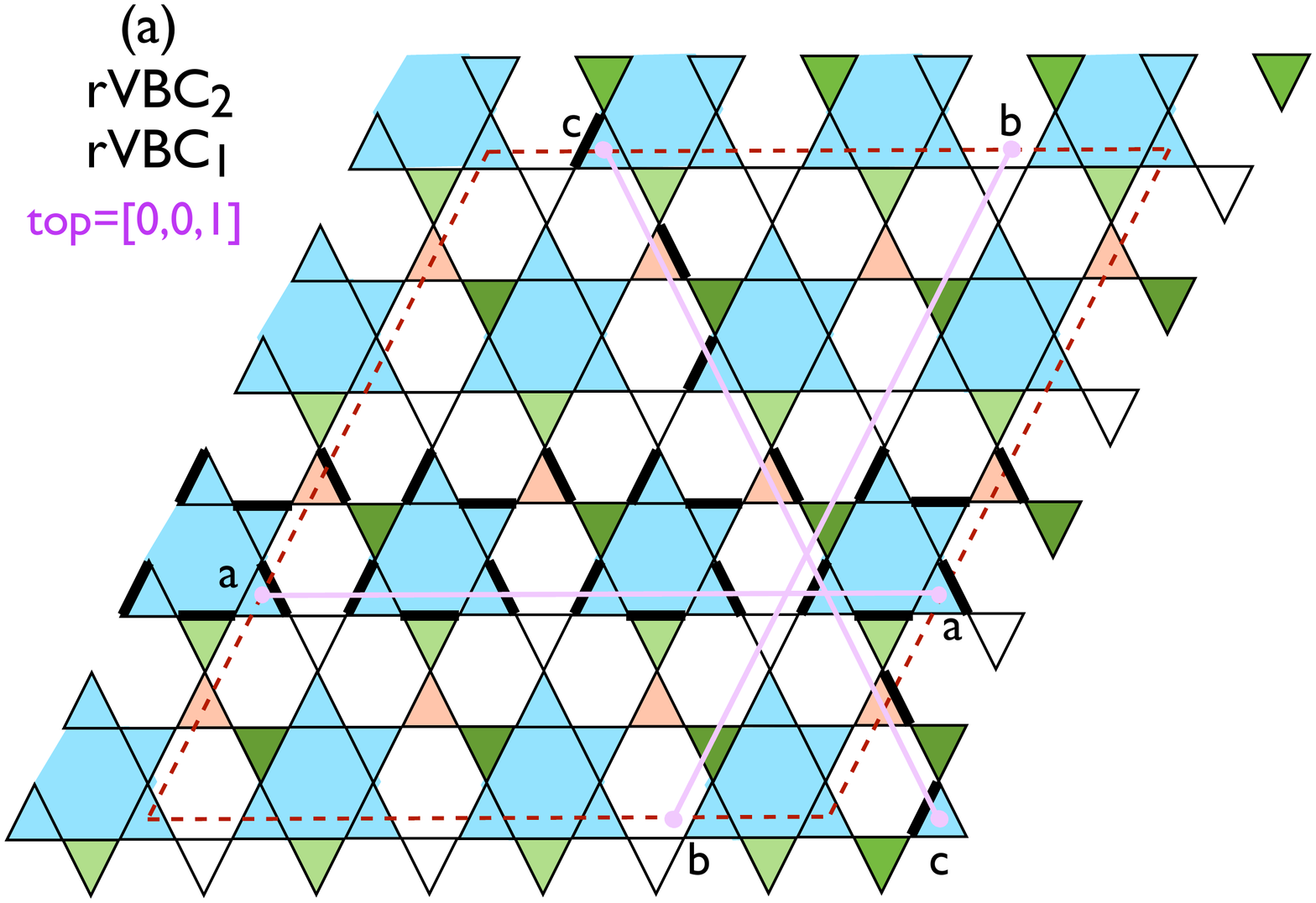}
\vskip -0.6cm
\includegraphics[width=0.9\columnwidth]{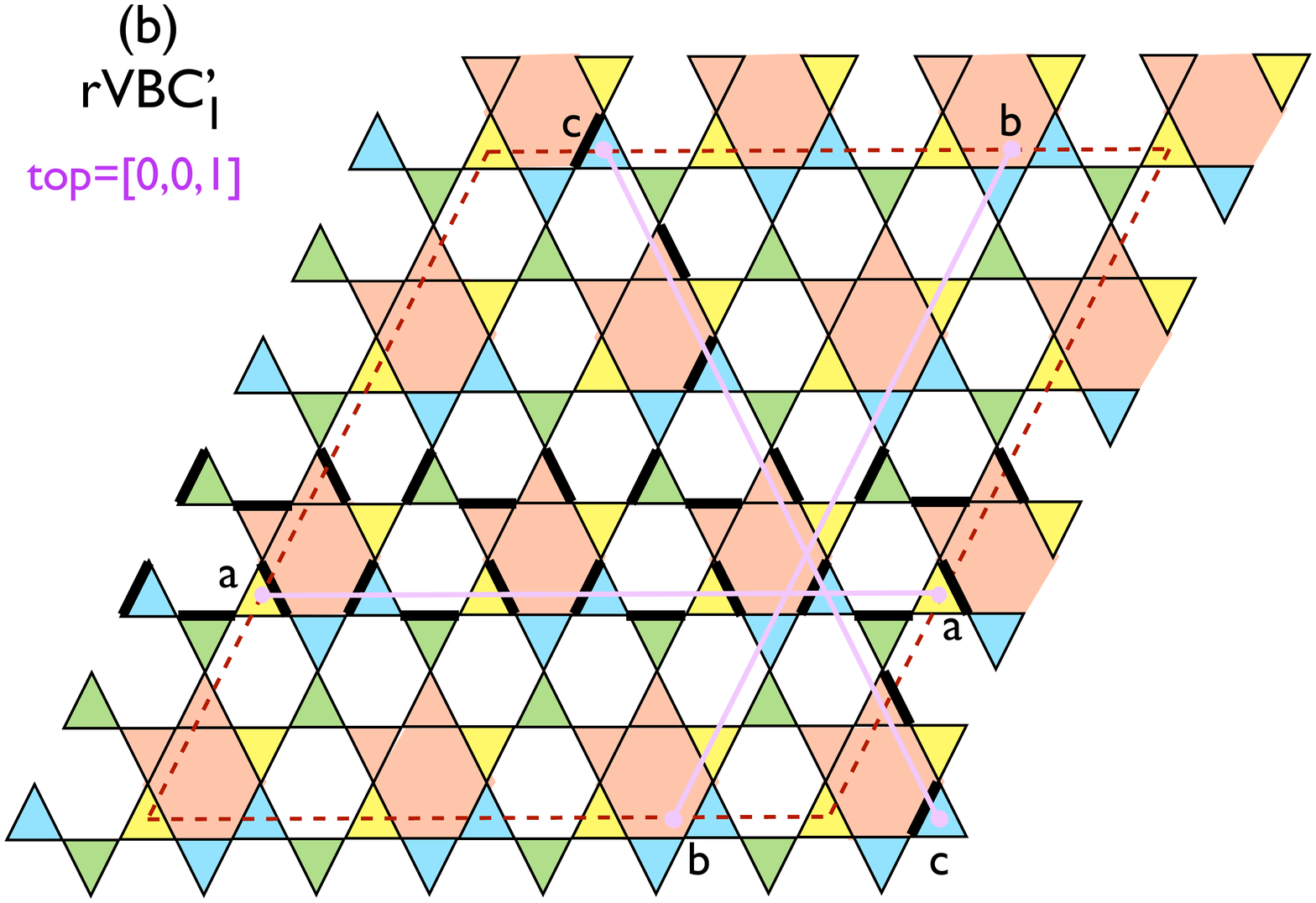}
\vskip -0.6cm
\includegraphics[width=0.9\columnwidth]{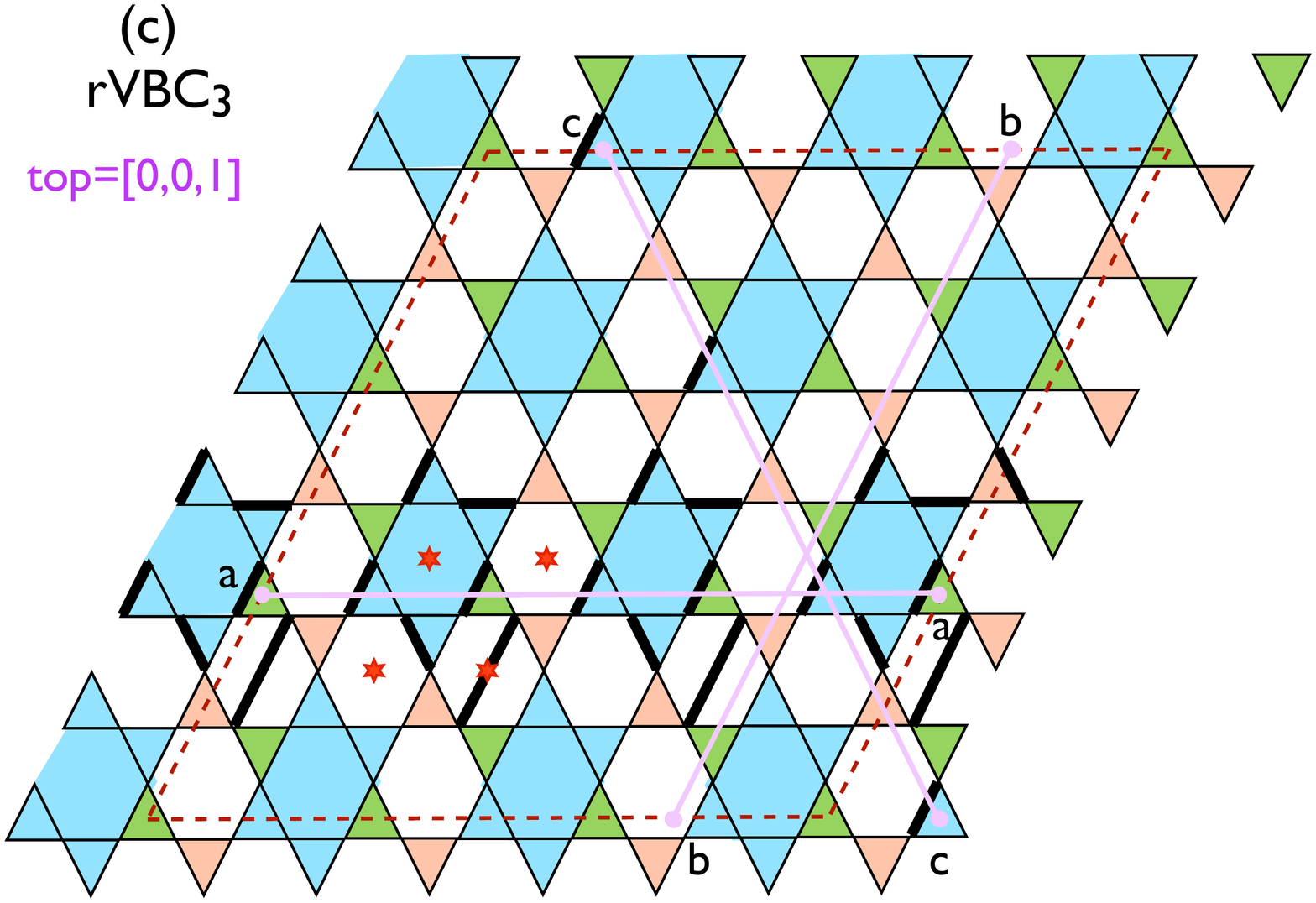}
\vskip -0.6cm
\caption{Resonant VBC with $2\times 2$ supercells in the [0,0,1] topological sector of the periodic $N=108$ site cluster.
Resonating 10-dimer (a,c) and 8-dimer (b) loops (see text) are represented by shaded areas. 
(a) When $\alpha=1$ (see text) a symmetry axis perpendicular to the c direction is restored (light and dark green triangles become
equivalent).
(b) The pattern exhibits a reflection symmetry w.r.t. the c direction.
(c) Inversion symmetry w.r.t. the centers of all the hexagons is present.
Dimers beyond NN have been used. }
\label{fig:kagome3}
\end{figure}

We have computed these multiplicities for a few VBC with a $2\times 2$ or $2\times 1$ supercell on the 108-site
cluster.
These multiplicities were obtained in a fixed topological sector, in order to compare to the LED results (next section). To do so, the lattice symmetry group was simply restricted to the elements which conserve the topological sector, {\it i.e} excluding $2\pi/3$ rotations in the $[0,0,1]$, $[0,1,0]$ and $[1,0,0]$ sectors. The results are shown in Table~\ref{table:VBC}. The VBCs named CVBC is the simplest, since, as far as the symmetries are concerned, it can be reduced to a single dimer covering.
CVBC is displayed in Fig.~\ref{fig:kagome1}(b), and has first been proposed in Ref.~\onlinecite{ba04}. It has a $2\times 1$ supercell and a degeneracy equal to 8 (for each of the 3 topological sectors, so 24 in total).  
The VBC$_0$ of Fig.~\ref{fig:kagome2}(a) can also be reduced to a single dimer covering but, in contrast to the CVBC,
belongs to the [1,1,1] topological sector. Its degeneracy is 24.
The other VBC we considered involve some resonance between different dimer configurations. The $r{\rm VBC}_2$ 
(see Fig.~\ref{fig:kagome3}(a)) can be constructed from the CVBC configuration by applying some resonance operators $\sigma^x$ (notation of Ref. \onlinecite{msp02})
on a quarter of the hexagons
(colored in blue in Fig.~\ref{fig:kagome3}(a)),
to shape a $2\times 2$ supercell: 
$$|r{\rm VBC}_2\rangle=\prod_{h=1,...,9} (1+\alpha\sigma^x_h) |{\rm CVBC}\rangle.$$ 
Doing so simultaneously enlarges the supercell and doubles
the degeneracy (16). If the coefficient $\alpha$ is set to one, one restores a reflection symmetry ($\sigma=+$) and we get another VBC, named  $|r{\rm VBC}_1\rangle$ and with degeneracy $8$.
By applying the resonance operators on the other hexagon-sublattice
(colored in orange in Fig.~\ref{fig:kagome3}(b)),
one obtains another resonating VBC, 
$|r{\rm VBC}_1^\prime\rangle$ shown in Fig.~\ref{fig:kagome3}(b),
with the same degeneracy (8) but different quantum numbers (the symmetry axis is now that of the wave vector B$^*$ i.e.
along the c direction).
Another resonating crystal, $|r{\rm VBC}_3\rangle$ shown in Fig.~\ref{fig:kagome3}(c), with no reflection symmetry (in contrast to the previous ones) but with inversion symmetry w.r.t the centers of all the hexagons can also be constructed in the [0,0,1] topological sector but using longer range dimers. We believe there is a similar VBC (that we name VBC$^*$) which bears exactly the same symmetry property and which could be 
represented solely in terms of (hardcore) NN dimers, probably requiring a more involved resonance structure.
Finally, the SVBC  discussed in Ref.~\onlinecite{sm02} is made of resonating ``stars'' (twelve-dimer loops) and also has a $2\times 2$ supercell. Similarly, the DVBC is made of resonating diamonds (eight-dimer loops).
But unlike the previous examples, these two crystals belong to the [1,1,1] sector.

It appears that the trivial IRREP ($\Gamma(+,+,+)$ or $\Gamma^*(\times,+,+)$) appears {\it exactly once} in all the VBC we have listed so far.
This is in fact a general property of such kind of broken symmetry phase. Indeed, the multiplicity of $\Gamma$ is just a special case of Eq.~\ref{eq:mult}
where the character $\chi_\Gamma$ is constant and equal to unity. We thus have
$$
n_\Gamma=\frac{1}{|G|}\sum_{g\in G} {\rm Tr}_{|1\rangle,\cdots |d\rangle}(g).
$$
However ${\rm Tr}_{|1\rangle,\cdots |d\rangle}(g)$ is nothing but the number of states which are invariant under the symmetry $g$. So $n_\Gamma$ can be written using the stabilizers
${\rm Stab}(i)$ or each state $|i\rangle$:
$$
n_\Gamma=\frac{1}{|G|}\sum_{i=1}^d | {\rm Stab}(i)|.
$$
But since all the stabilizers have the same number of elements, $|{\rm Stab}(i)|=|G|/d$, we just find $n_\Gamma=1$.
In Sec.~VI this remark will allow us to show that a particular multiplet of low-lying states cannot be interpreted as a single conventional VBC.

\section{Numerical results}

\begin{figure}[htb]
\includegraphics[width=0.9\columnwidth]{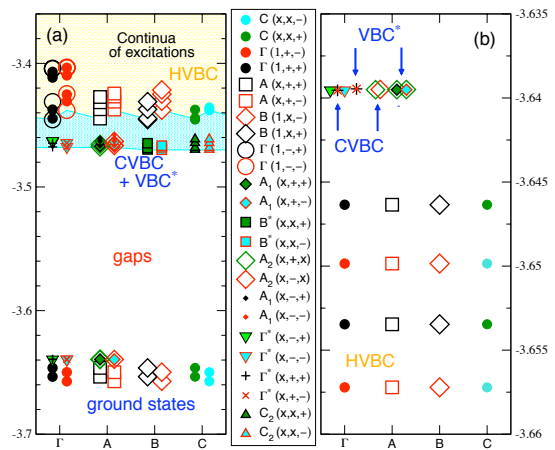}
\caption{
(a) Low energy spectrum of the 108-site cluster in the  $T_0=[1,1,1]$ 
and $T_3=[0,0,1]$ (degenerate with $[0,1,0]$ and $[1,0,0]$) topological sectors.
Different symbols distinguish different momenta and topological sectors (see Fig.~\ref{fig:BZ}(a) \& (b)) 
and IRREPs labeled as "$(r_3,r_2,\sigma)$" (see text). Only a few levels corresponding to the on-sets
of the two continua (shaded areas) are shown. (b) Zoom in of the (quasi-) ground states. 
An Ising degeneracy (see text) is reflected by the (almost) perfect equal spacings
between the HVBC ground states. }
\label{fig:spectra}
\end{figure}

We have used the Lanczos Exact Diagonalization (LED) technique to obtain the low energy spectra
of the QDM Hamiltonian in each IRREP. Computations have been performed on the Altix SGI cluster at 
CALMIP (Toulouse), 
in particular on the Altix UV node providing 96 CPUs allocating 1024 Gb of fast access RAM.
Typically, only 200 Lanczos iterations were sufficient to get at least 5 or 6 of the lowest eigenenergies (in each IRREP)
with full machine (double) precision.
Fig.~\ref{fig:spectra}(a) shows the obtained low-energy spectra at the most relevant momenta and in
the two topological sectors [1,1,1] and [0,0,1]. For each of these sectors, we can see
a group of quasi-degenerate levels named "ground state manifold" and a dense accumulation of levels at higher energy.
This clearly reflects the presence of two continua of excited states above two (slightly different) thresholds 
separated from the GS manifolds by two gaps. Note that the energy 
spectra at the other momenta (D, E and F in Fig.~\ref{fig:BZ}) are not shown on this picture
since they only contain excited states beyond the gaps and no level belonging to the ground state manifold. 
Fig.~\ref{fig:spectra}(b) shows a zoom in of the ground state manifold which reveals two subgroups
corresponding to each topological sector.

In agreement with Ref.~\onlinecite{poilblanc}, the quantum numbers of the GS levels 
in the "symmetric" [1,1,1] topological sector correspond exactly to those predicted (see Ref.~\onlinecite{sm07} for details) 
for the 36-site HCVB schematically shown in Fig.~\ref{fig:kagome1}.  In addition, we find here some very remarkable 
(Ising-like) fine structure with equally spaced sub-subgroups of levels which shall be discussed later on.
The numerical values of these eigenenergies reported in Table~\ref{table:energy1} reveal
almost exact degeneracies within less than $10^{-5}$ (in units of the exchange constant $J$). 

\begin{table}[tbh]
\begin{center}
 \begin{tabular}{@{} cccccc @{}}
\toprule
IRREP& $\#$ of & $\#$ of   & $\#$ of & GS  & total \\
& $\bf K$ & states  &  levels$^\dagger$  & energy levels & degen.$^\ddagger$ \\
\hline
$\Gamma\, (1,+,-)$ & 1 &  $79\, 548\, 096$ & 2 & $-3.657\, 214\, 41$ & 2 \\
           &  &  &  & $-3.649\, 849\, 41$ &  \\
$\Gamma\, (1,+,+)$ & 1 &  $79\, 548\, 096$ & 2 & $-3.653\, 462\, 67$ & 2 \\
           & &  &  & $-3.646\, 352\, 04$ &  \\
$A\, (\times,+,-)$ & 3 &  $238\, 642\, 176$ & 2 & $-3.657\, 214\, 51$ & 6 \\
          &  &  &  & $-3.649\, 849\, 60$ &  \\
$A\, (\times,+,+)$ & 3 &$238\, 642\, 176$ & 2 & $-3.653\, 462\, 48$ & 6 \\
           &  &  &  & $-3.646\, 351\, 92$ &  \\
$B\, (1,\times,-)$ & 2 &  $159\, 073\, 536$ & 2 & $-3.657\, 213\, 37$ & 4 \\
           &  &  &  & $-3.649\, 849\, 46$ &  \\
$B\, (1,\times,+)$ & 2 &  $159\, 073\, 536$ & 2 & $-3.653\, 462\, 55$ & 4 \\
          &  &  &  & $-3.646\, 352\, 03$ &  \\
$C\, (\times,\times,-)$ & 6 &  $477\, 218\, 560$ & 2  & $-3.657\, 213\, 33$ &  12\\
           &  &  &  & $-3.649\, 849\, 58$ &  \\
$C\, (\times,\times,+)$ & 6 &  $477\, 218\, 560$ & 2 & $-3.653\, 462\, 56$ & 12 \\
         &  &  &  & $-3.646\, 351\, 92$ &  \\
\toprule
\end{tabular}
\end{center}
\caption{Quasi-degenerate GS classified according to the IRREP of the
$C_{6v}\otimes {\cal T}$ space group of the $[1,1,1]$ topological sector of the 108 cluster. 
$^\dagger$The Lanczos algorithm does not (easily) provide the exact degeneracy of each energy level. 
$^\ddagger$An extra (exact) 
degeneracy of 2 is expected for each level (see text).}
\label{table:energy1}
\end{table}

The investigation of the [0,0,1] and the two other equivalent topological sectors was overlooked in Ref.~\onlinecite{poilblanc}.
Interestingly, Fig.~\ref{fig:spectra}(b) reveals a new subgroup of energy levels within $\sim 6\times10^{-3}J$ from
the [1,1,1] GS subgroup. Their energies are listed in Table~\ref{table:energy2} showing again almost perfect degeneracies
within sub-subgroups of levels. 
The quantum numbers of these states can be straightforwardly compared to the expected quantum numbers for
the VBC shown in Table~\ref{table:VBC}. Surprisingly, one finds that the numerical results do not correspond 
to any {\it single} VBC. However, they correspond exactly to the {\it reunion} of the GS multiplets of CVBC and VBC$^*$. 

\begin{table}[tbh]
 \begin{tabular}{@{} cccccc @{}}
\toprule
IRREP& $\#$ of & $\#$ of   & $\#$ of & GS  & total \\
& $\bf K$ & states  &  levels$^\dagger$  & energy levels & degen.$^\ddagger$ \\
\hline 
$\Gamma^*(\times,+,+)$ & 1 &  $238\, 613\, 056$ & 2 & $-3.639\, 546\, 99$  & 2 \\
  &   &  &  & $-3.639\, 460\, 27$  & \\
$\Gamma^*(\times,+,-)$ & 1 &  $238\, 613\, 056$ & 2 & $-3.639\, 546\, 42$ & 2 \\
  &   &  &  & $-3.639\, 458\, 83$  & \\
$\Gamma^*(\times,-,+)$ & 1 &  $238\, 613\, 056$ & 1 & $-3.639\, 497\, 82$ & 1 \\
$\Gamma^*(\times,-,-)$ & 1 &  $238\, 613\, 056$ & 1 & $-3.639\, 
495\, 81$ & 1 \\
$A_1\, (\times,+,+)$    & 1 &  $238\, 605\, 760$ & 1 &  $-3.639\, 509\, 03$ &  1 \\
$A_1\, (\times,+,-)$     & 1 &  $238\, 605\, 760$ & 1 &   $-3.639\, 509\, 00$      & 1\\
$A_2\, (\times,+,\times)$    & 2 &  $477\, 218\, 816$ & 3 &  $-3.639\, 508\, 79$  &  6 \\
  &   &  &  & $-3.639\, 508\, 60$  & \\
  &   &  &  & $-3.639\, 497\, 19$  & \\
$A_2\, (\times,-,\times)$     & 2 &  $477\, 218\, 816$ & 1 &  $-3.639\, 496\, 79$  & 2\\
\toprule
\end{tabular}
\caption{
Quasi-degenerate GS classified according to the IRREP of the
$C_{2v}\otimes {\cal T}$ space group of the $[0,0,1]$ topological sector of the 108 cluster. 
$^\dagger$The lanczos algorithm does not provide (easily) the exact degeneracy of each energy level.}
\label{table:energy2}
\end{table}

\section{Discussions}

Our results clearly show a close competition between two (or even three) types of VBCs. We have checked 
that very small changes of
the Hamiltonian parameters can easily reverse the relative position of the two subgroup of levels. Therefore it is not clear
which of these VBC will be stabilized in the thermodynamic limit. 

Let us first discuss the remarkable fine structure of the GS manifold in the [1,1,1] topological sector.
It reflects the hidden Ising variable~\cite{singh} associated to the two possible chiralities, 
let's say $\uparrow$ and $\downarrow$, of each
of the resonating stars (or pinwheels) of the HVBC. On our 108-site cluster 3 such resonating stars can be fitted 
leading to $2^3=8$ possible configurations. The numerical spectrum can be explained by assuming a 
(small) effective "magnetic" Ising field $h$ which naturally provides splitting by energies $-3h$, $-h$, $h$ and $3h$.
Although this structure is relatively robust and does not require real "fine tuning" of the effective model,
the fact that $J_{12}=0$ is important. Indeed, when $J_{12}$ is added by hand, the numerical results can be simply interpreted
by assuming a (strong) linear dependance of the effective Ising field with $J_{12}$, $h=h_0+ A J_{12}$, and $h$ changes sign for a tiny (positive) value of $J_{12}\sim 0.001$.
Similarly, a small value of $J_{12}=\pm 0.01$ increases the splitting within the GS manifold 
by a factor $\sim 5$. The behavior in the thermodynamic limit (for $J_{12}=0$) depends crucially whether the effective 
$h_0$ is exactly zero 
or retains a very small but still finite value. In the latter (most probable) case, the absolute 
HVBC GS should exhibit a uniform 
chirality of all resonating stars (one of the two ferromagnetic Ising configurations) and the VBC gap should be filled 
by a discrete set of energy levels corresponding to the excitations of a finite number of resonating stars. 
In contrast, in the unlikely case where $h_0=0$ (due to fine-tuning) 
the HVBC GS would retain an infinite degeneracy.

Identifying the exact nature of the phase in the $T_1$, $T_2$ and $T_3$ topological sectors
is more subtle. In fact, because of two (GS) energy levels in the fully 
symmetric $\Gamma^*(\times,+,+)$ IRREP
this phase can not be described as a unique VBC.  From a direct comparison of the quantum numbers
of the states in the GS manifold with those listed in Table~\ref{table:VBC}, our numerical data
can indeed be interpreted as the simultaneous occurrence of two very closely competing VBC, the columnar CVBC and the $2\times 2$ VBC$^*$. 
Such a description however raises a number of concerns. 
First, the exact nature of VBC$^*$ is not fully known: it is similar to SVBC in terms of symmetry properties but belongs to a different topological sector. A naive representation in terms of a simple
NN dimer covering is therefore impossible and should involve subtle resonances like
rVBC1 or rVBC2. 
Secondly, we have checked that the almost exact degeneracy between CVBC and VBC$^*$ is 
robust and survives under small changes of the QDM Hamiltonian parameters so that it does not
correspond to any "fine tunning".  
This unconventional feature might have different origins:
(i) CVBC and VBC$^*$ are truly different but are related by some hidden symmetry of the
original projected Heisenberg model or of its approximate effective model (\ref{eq:Hefffull});
(ii) The relevant phase still corresponds to a unique VBC of smaller degeneracy ($3\times 8=24$
instead of $3\times 16=48$) but admits some form of zero-energy excitations. 
It is interesting to note that (\ref{eq:Hefffull}) only contains loops of different lengths 
but always involving a {\it single} hexagon. It might be that smaller terms involving longer loops
(and hence at least two hexagons) could lift this degeneracy.
(iii) The observed degeneracy is the signature of an ordered phase which is more complex than a VBC.
We could think, for instance, of some unconventional VBC  with broken time-reversal symmetry. Although
it is not clear to us how to construct precisely such a state, it may have $n_\Gamma=2$.
The tight competition between several VBC in addition to the existence of a
$\mathbb{Z}_2$ dimer liquid stable very nearby in parameter space~\cite{poilblanc}
are clear signatures of the highly frustrated nature of the generalised QDM model,
probably also generic in several highly frustrated QHAF. 
In fact, the observation in recent DMRG studies of the kagome QHAF~\cite{steve}
of signals of the formation of crystalline diamond patterns 
together with finger prints of a $\mathbb{Z}_2$ {\it spin} liquid suggest
that the QHAF and the effective QDM might have strong similarities. 
In any case, our results reveal an exceptionally large quasi-degeneracy of the GS multiplets
($96+48=144$ on our 108-site cluster !) which might reflect the proximity of the $\mathbb{Z}_2$ 
dimer liquid.\cite{poilblanc} 
If the same were to exist in a frustrated QHAF, we believe the related LED and DMRG results might become
quite difficult to analyse. Incidentally, our study also shows that the existence of 
quasi-degenerate GS in different topological sectors (on the cylinder or on the torus) 
is not sufficient to completely identify a $\mathbb{Z}_2$ dimer (or spin) liquid, our result providing
a counter-example. 

\section{Acknowledgements} 

DP acknowledges support by
the French Research Council (Agence Nationale de la
Recherche) under grant No. ANR 2010 BLANC 0406-01. 
DP is also grateful to Nicolas Renon at CALMIP
(Toulouse, France) and to SGI (France) for support in the use of the Altix
SGI supercomputer.
DP thanks Matthieu Mambrini (LPT) and Arnaud Ralko (Institut N\'eel, Grenoble)
for numerous discussions and comments. 



\begin{thebibliography}{99}
\bibitem{note-rescale}  A rescaling of the QHAF hamiltonian is performed
$\hat{\cal H}_H\rightarrow (4/3)\hat{\cal H}_H+JN/2$.
\end{thebibliography}
\end{document}